\documentclass{ifacconf_amx}

\usepackage{graphicx}      
\usepackage{float}  

\usepackage{subcaption} 
\usepackage{tabularx}      
\usepackage{makecell}      
\usepackage{multirow}      
\usepackage{amsmath,amssymb,amsfonts} 
\usepackage{xcolor}        
\usepackage{cite}          

\usepackage{siunitx}       
\usepackage{fancyhdr}      

\usepackage{lipsum}        

\usepackage[scaled]{helvet}
\usepackage[T1]{fontenc}

\newcommand{\ISBN}{UNKNOWN ISBN}

\newcommand{\PaperTitle}{A title for your paper}
\newcommand{\PaperDate}{01 April 1900}
\newcommand{\AuthorFooter}{Mustermann, M.,Supervisor, S., Oksanen T.}

\makeatletter
\let\old@ssect\@ssect 
\makeatother

\usepackage[hidelinks]{hyperref}      
\usepackage{cleveref}      

\makeatletter
\def\@ssect#1#2#3#4#5#6{%
	\NR@gettitle{#6}
	\old@ssect{#1}{#2}{#3}{#4}{#5}{#6}
}
\makeatother

\renewcommand{\ISBN}{978-3-911430-12-8}
\renewcommand{\PaperTitle}{EmbC-Test: How to Speed Up Embedded Software Testing Using LLMs and RAG}
\renewcommand{\PaperDate}{09 March 2026}
\renewcommand{\AuthorFooter}{Harnot, M., Komarnicki, S., Polok, M., Oksanen T.}
\begin{document}
\begin{frontmatter}

   \title{\PaperTitle}

   \author[Second]{Maximilian Harnot}
   \author[Second]{Sebastian Komarnicki}
   \author[Second]{Michał Polok}
   \author[First]{Timo Oksanen}

   \address[First]{Technical University of Munich, Germany \\ Professorship of Agrimechatronics; and \\ Munich Institute of Robotics and Machine Intelligence (MIRMI)}
   \address[Second]{Hydac Software GmbH, Germany}

   \begin{abstract}
Manual development of automatic tests for embedded C software is a strenuous and time-consuming task that does not scale well. With the accelerating pace of software release cycles, verification increasingly becomes the bottleneck in the embedded development workflow. This paper presents a Retrieval-Augmented Generation (RAG) pipeline as a solution for partial automation of the verification process. By grounding a large language model in project-specific artifacts, the approach reduces hallucinations and improves project alignment. An industrial evaluation showed that the generated tests are 100 \% syntactically correct, with 85\% successfully passing runtime validation. The proposed solution has the potential to save up to 66\% of the testing time compared to manual test writing while generating 270 tests per hour.
        
   \end{abstract}


\end{frontmatter}

\section{Introduction}

The pace of embedded software development is accelerating, driven by more frequent releases and the increasing use of AI for code generation. While implementation throughput rises, verification effort, especially in safety-relevant embedded domains where evidence, traceability, and reproducibility are necessary, scales poorly. Manual implementation of unit tests for embedded C code consumes scarce testing resources and struggles to keep up with shorter release cycles. But naïve LLM-generated tests are not just noisy, they can be wrong in subtle ways and create false confidence.

At first glance, LLMs appear to offer a compelling shortcut: generate new automatic tests, increase coverage, and reduce time-to-validation. However, such an approach, known as "zero-shot" test generation, is unreliable. Without deep project context, LLMs frequently misuse internal APIs, invent functions or types, and deviate from established test conventions and house style. More critically, even when generated tests compile and execute, they can encode incorrect expectations or introduce flakiness via implicit dependencies and incomplete mocking. In safety-critical environments, such failures create false confidence.

This paper, based on the results obtained in \cite{Harnot2026}, presents an approach to making LLM-based test generation more trustworthy: Retrieval-Augmented Generation (RAG) combined with a quality assurance pipeline. Instead of relying solely on an LLM’s general training, RAG grounds generation in the organization’s own engineering artifacts, such as requirements, existing tests, C source and header files, and documentation. The central hypothesis is that project-aware retrieval can reduce hallucinations, improve adherence to internal patterns, and enable systematic expansion of test suites with boundary-condition and fault-handling scenarios that are essential in embedded control software.

\begin{figure}
   \centering
   \includegraphics[width=0.85\linewidth]{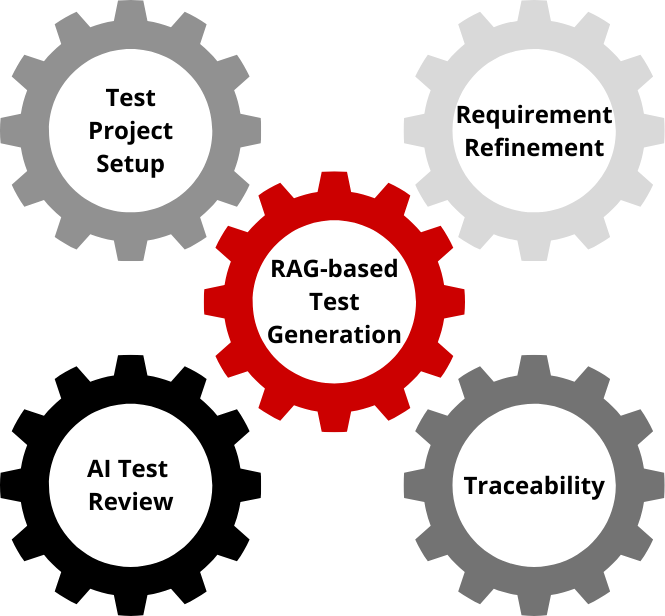}
   \caption{AI-assisted software testing ecosystem at Hydac Software}
   \label{fig:ecosystem}
\end{figure}

At Hydac Software, the testing workflow is supported by a comprehensive toolchain designed to ensure quality and traceability. This ecosystem begins with the automated creation of a test project tailored to the unit being tested. Once tests are created and executed, an AI test review tool analyzes logs against requirements producing a human-readable report, and a traceability tool ensures complete requirement to test mapping. Finally, a requirement refinement tool analyzes and improves requirements, identifying gaps in test coverage and suggesting additional requirements. Within this ecosystem, the actual creation of test cases remained a manual, time-consuming bottleneck. The RAG-based test generator presented in this work acts as the crucial missing piece. By automating test creation, it completes an end-to-end AI-assisted testing pipeline (Figure~\ref{fig:ecosystem}).

A key message of this work is that scaling tests is not the same as scaling confidence. Code coverage is an important indicator of under-tested regions, yet it must be interpreted carefully and complemented with stronger measures of effectiveness and oracle quality.

Consequently, this paper emphasizes multi-dimensional evaluation: retrieval quality metrics, automated validity checks, and additional quality measures beyond coverage where feasible.

The contribution of this work is twofold. 

First, it provides a pragmatic reference architecture for a RAG-powered test generation pipeline (knowledge base construction, chunking strategies, embeddings, hybrid retrieval, prompting, and integration with a selected LLM platform).

Second, it proposes an evaluation approach tailored to embedded contexts, including baselines (random chunk retrieval, non-RAG generation) and practical system metrics such as latency and throughput. 
The overall goal is not to replace human expertise but to shift expert time toward high-leverage activities, like risk analysis, exploratory testing, and oracle refinement. The contributions of the work go far beyond theoretical research as the resulting tool has been deployed and is actively used in the Hydac Software testing workflow.

\section{Background}

Reviewing the existing methods of automated test generation revealed a significant gap. While classical tools, including formal verification, search-based generation, and random testing frameworks \cite{CBMC2023,KLEE2008,EvoSuite2011,Randoop2007} are well-established, they are primarily focused on high-level programming languages. Moreover, existing solutions rarely incorporate project-specific documentation or legacy tests into the generation process.

LLMs have emerged as a shortcut to faster test generation and elimination of some of these limitations. Solutions like ChatUniTest \cite{Chen2024} have proven that iterative prompting and project-awareness outperform zero-shot generation \cite{Codex2021,Wang2024}. Nevertheless, issues like hallucinating API calls, generating flaky assertions, or incomplete dependencies persist regardless of the model choice.

A promising solution to these challenges is Retrieval-Augmented Generation (RAG), which grounds the LLM in project-specific artifacts, diminishing the hallucination problem and improving assertion quality \cite{Barnett2024}. Similar to previous methods, research has primarily focused on high-level programming languages, while embedded C remains underexplored in the context of AI-assisted and context-aware test generation. This work addresses that verification gap by combining the scalability of LLM-based test generation with project-specific context, enabling automated test generation for embedded C software.

\section{Problem Statement and Objectives}
The core challenge of the project is to automate Python-based test generation for embedded C with the aim of:
\begin{itemize}
    \item Produce executable tests with correct syntax that compile and run against the target software
    \item Align with the project APIs, data structures and coding conventions while reducing hallucinations
    \item Improve code coverage of critical branches with meaningful assertions
    \item Achieve a throughput that significantly outpaces manual test creation
    \item Reduce the repetitive task of test creation and refinement, substituting it with short review and correction sessions of the generated tests
\end{itemize}

In order to meet these goals, the following design and implementation objectives were established:
\begin{itemize}
    \item Constructing a knowledge base of C headers, source code, and legacy tests that are chunked, vectorized, and prepared for retrieval
    \item Implementing hybrid retrieval (dense embeddings + BM25) that captures both syntactic and semantic similarity
    \item Designing a RAG pipeline that automates test generation by incorporating retrieved chunks into a prompt containing project specifications, documentation, and legacy test templates
    \item Creating an evaluation framework to assess the performance of the pipeline across five dimensions: code coverage, test correctness, retrieval quality, computational efficiency and human evaluation
    \item Analyzing the results in order to find the best performing configuration and evaluate its benefits for industrial applications
\end{itemize}

Meeting these objectives will result in a shift of focus from repetitive test development to reviewing, improving, and refining the generated tests, which will lead to the acceleration of the verification cycle without compromising test quality.

\section{System Architecture}
The system is subdivided into two phases. The offline phase chunks the project artifacts into retrievable units, which are embedded into vector representations and stored as indices, ready for retrieval by both semantic and lexical search. The online phase takes the contents of a single software requirement, matches it against the indices and ingests the most relevant documentation and legacy test sections into the prompt, which is then sent to a cloud-hosted LLM. The tests generated by the LLM are then received and saved.

\subsection{Knowledge Base and Chunking Strategies}

The project includes a knowledge base containing C header files, source code, and legacy Python tests.

Choosing an appropriate chunking strategy is essential, as it improves the retrieval quality and helps to capture relevant context. 

\pagebreak

Three chunking strategies were implemented for the embedded C code and one for the Python tests:

\begin{itemize}
    \item \textbf{Fixed-size chunking}: Dividing code into segments of fixed size. This is a very basic chunking method that runs the risk of splitting functions.
    \item \textbf{Brace-aware chunking}: Dividing code by counting brackets and splitting chunks at zero-depth points. It guarantees that functions and structures are not split apart. 
    \item \textbf{AST-based chunking}: Parsing the source code into an Abstract Syntax Tree (AST) to map its underlying syntactic structure. Structures, type definitions, functions, enums etc. are identified and grouped into semantically meaningful chunks that preserve the logical structure of the code.
    \item \textbf{Test unit chunking}: Dividing test code by single test units.
\end{itemize}

After chunking, each chunk is transformed into a vector embedding using a locally hosted embedding model and stored in the vector database to enable semantic retrieval.

\begin{figure}[H]
   \centering
   \includegraphics[width=0.85\linewidth]{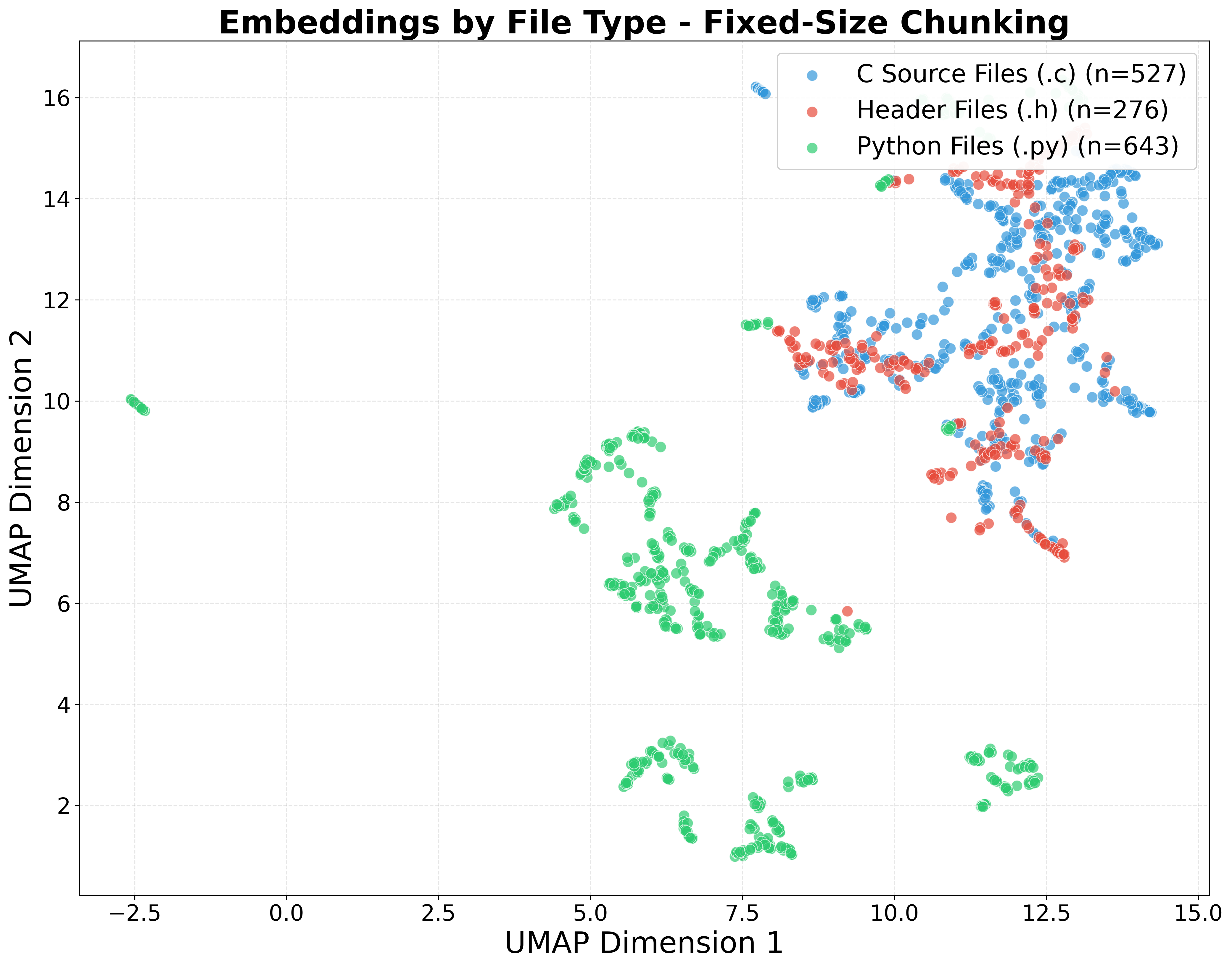}
   \caption{UMAP visualization of the embedding space for fixed-size chunking (1446 chunks). The clusters overlap due to arbitrary split boundaries.}
   \label{fig:umap-fixed}
\end{figure}

\begin{figure}[H]
   \centering
   \includegraphics[width=0.85\linewidth]{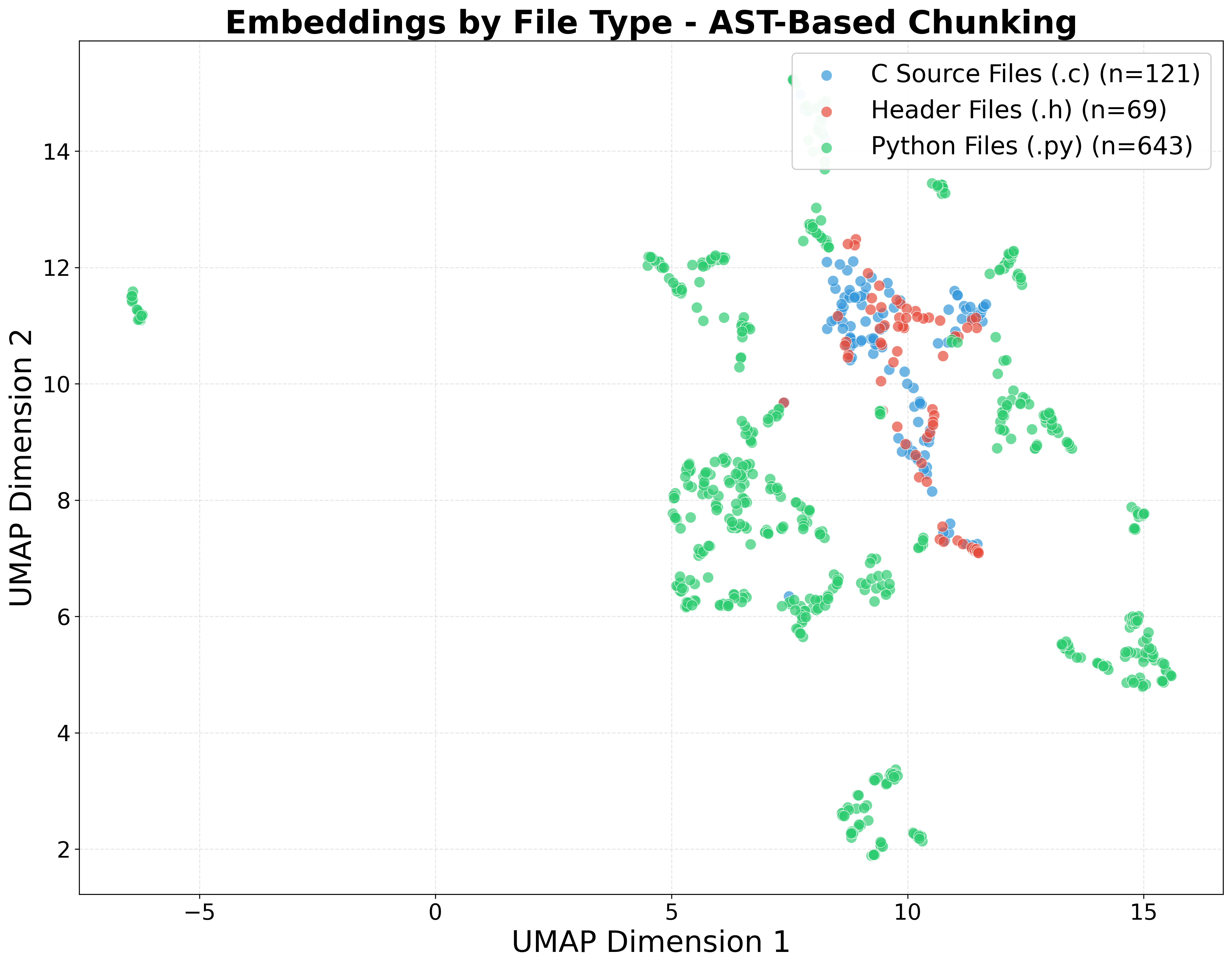}
   \caption{UMAP visualization of the embedding space for AST-based chunking (833 chunks). In AST-based chunking, the clusters are grouped more tightly together.}
   \label{fig:umap-ast}
\end{figure}

\subsection{Hybrid Retrieval}

The hybrid retrieval combines semantic and lexical search to retrieve the most relevant chunks from the vector database. It merges two methods that complement each other. Dense search uses cosine similarity to find the most similar vectors in the index database to the embedded query. This method is supported by the BM25 sparse search, which conducts a lexical matching with code-aware tokenization, looking for an exact word match, which may be missed by dense search. Both methods are merged by the Reciprocal Rank Fusion (RRF) function, which gives them equal weight. The top 5 results of the hybrid retrieval are then passed to the user prompt.

\subsection{Prompt Construction and LLM choice}

\begin{figure} [H]
   \centering
   \includegraphics[width=0.95\linewidth]{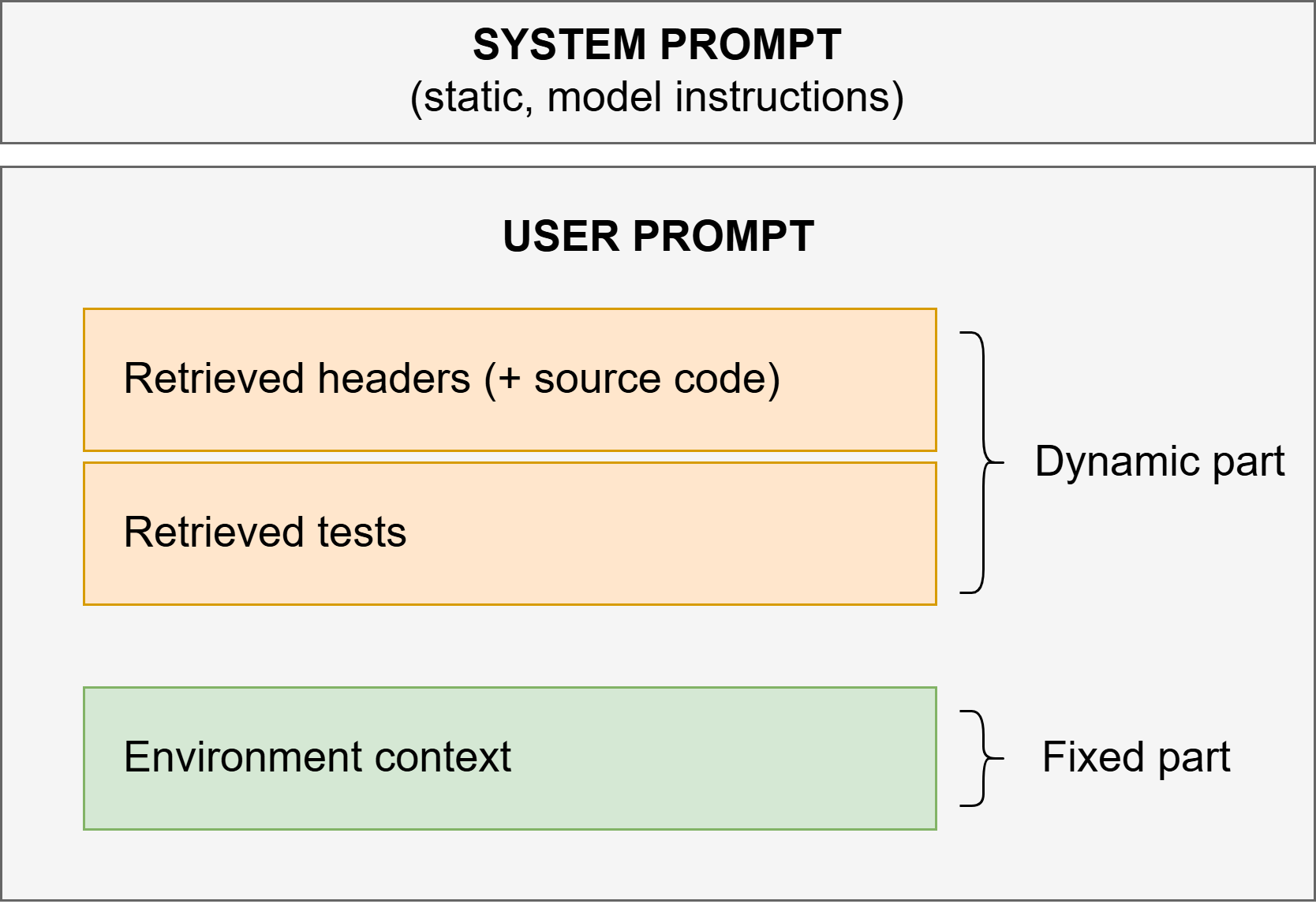}
   \caption{Test generation prompt structure.}
   \label{fig:prompt-structure}
\end{figure}

The prompt is designed to guide the LLM in generating syntactically correct and project-aligned test cases by providing structured instructions and relevant contextual information. The test generation prompt consists of two parts: a system prompt and a user prompt (Figure~\ref{fig:prompt-structure}). The system prompt defines the behavior of the model, its constraints and the output format. In the user prompt, the first part is dynamic and includes the retrieved headers and source code snippets as well as the retrieved tests. Following that, a fixed part informs about the environment context and project-specific details, followed by the software requirement. That order ensures that the LLM receives the full context before analyzing the requirement.

\section{Evaluation Methodology}

The developed system was evaluated on embedded C code from Hydac Software GmbH. Combining two LLMs, two embedding models, four chunking methods, two knowledge bases, and two test libraries resulted in 64 different generation system configurations. From these, a representative subset was selected and supplemented with 2 random chunk configurations and 2 non-RAG baselines, resulting in a final test sample of 44 configurations.

\pagebreak

The evaluation was performed in five dimensions:
\begin{enumerate}
    \item \textbf{Code coverage}: Line and branch coverage was measured using standard C instrumentation.
    \item \textbf{Test correctness}: A three step validation of syntactic correctness, compilation (import) correctness, and runtime correctness.
    \item \textbf{Retrieval quality}: Precision@k and Recall@k were measured against a ground truth dataset for five software requirements. Precision@k indicates how many of the top retrieved chunks are relevant, while Recall@k measures how many of all relevant chunks are successfully retrieved.
    \item \textbf{System performance}: The latency and throughput were measured to assess computational efficiency.
    \item \textbf{Human evaluation}: Two independent domain experts used a five-point Likert scale to evaluate relevance, assertion correctness, edge-case completeness, and readability. The evaluation ends with a decision to accept, accept with modifications, or reject a given test.
\end{enumerate}

\section{Results}
Table~\ref{tab:correctness} shows the test correctness results, comparing RAG with the random chunk retrieval and no-retrieval baseline. Even though all three configurations show very high syntactic validation rates (only the no-retrieval baseline had some syntax errors), the runtime validation results clearly show that RAG achieves the highest scores.

\begin{table}[H]
    \centering
    \caption{Test correctness results including syntactic and runtime validation.}
    \label{tab:correctness}
    \begin{tabular}{l c c}
        \hline
        \textbf{Configuration} &
        \makecell{\textbf{Syntactic} \\ \textbf{Validation}} &
        \makecell{\textbf{Runtime} \\ \textbf{Validation}} \\
        \hline
        RAG                     & 100.0\% & 84.5\% \\
        Random retrieval (RND)  & 100.0\% & 62.4\% \\
        No retrieval (NOR)      &  96.8\% & 50.5\% \\
        \hline
    \end{tabular}
\end{table}

To supplement the correctness results, code coverage was also measured. The RAG-based system achieved up to 43\% branch coverage and 67\% line coverage. The existing manual test suites were benchmarked at 76\% branch coverage and 93\% line coverage. However, the manual test suite values were obtained after months of iterative refinement, while the RAG-based results were achieved in a single run, without any refinement or feedback.

\begin{figure}
   \centering
   \includegraphics[width=0.95\linewidth]{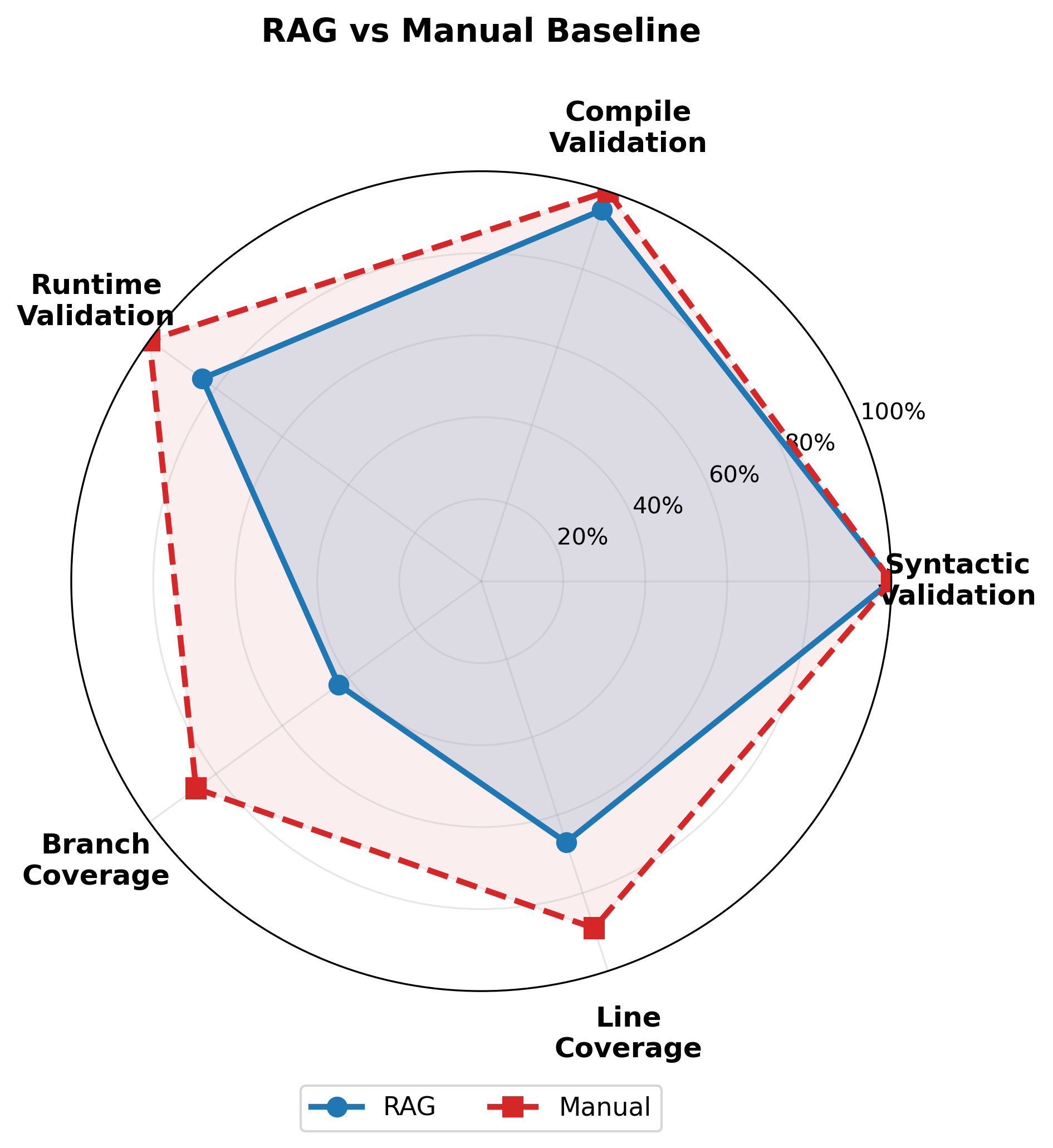}
   \caption{Comparison of the best-performing RAG configuration and manual test baseline.}
   \label{fig:radar-overview}
\end{figure}

A comparison of these parameters and how they correlate with each other can be seen in Figure~\ref{fig:radar-overview}.

\begin{figure}
   \centering
   \includegraphics[width=0.95\linewidth]{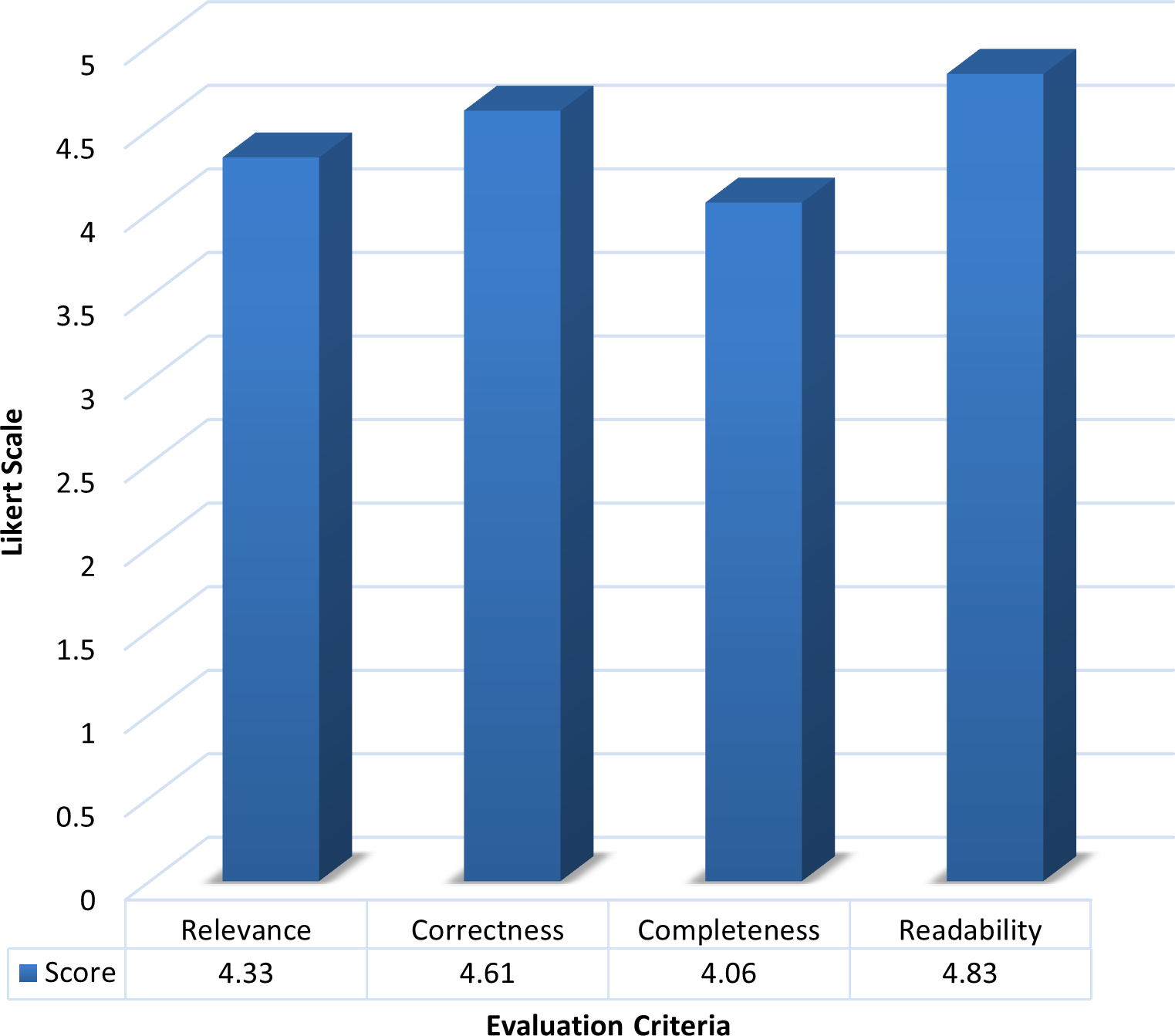}
   \caption{Likert scale scores for the best-performing RAG configuration.}
   \label{fig:likert-best}
\end{figure}

In addition to automatic evaluation, two independent domain experts assigned scores to the generated tests on a five-point Likert scale. The criteria included relevance, assertion correctness, edge-case completeness, and readability. The best-performing RAG configuration achieved 4.33 in relevance, 4.61 in correctness, 4.06 in completeness, and 4.83 in readability (Figure~\ref{fig:likert-best}), with a stunning \textbf{94.4\% test usability rate}. 

\begin{figure}
   \centering
   \includegraphics[width=0.95\linewidth]{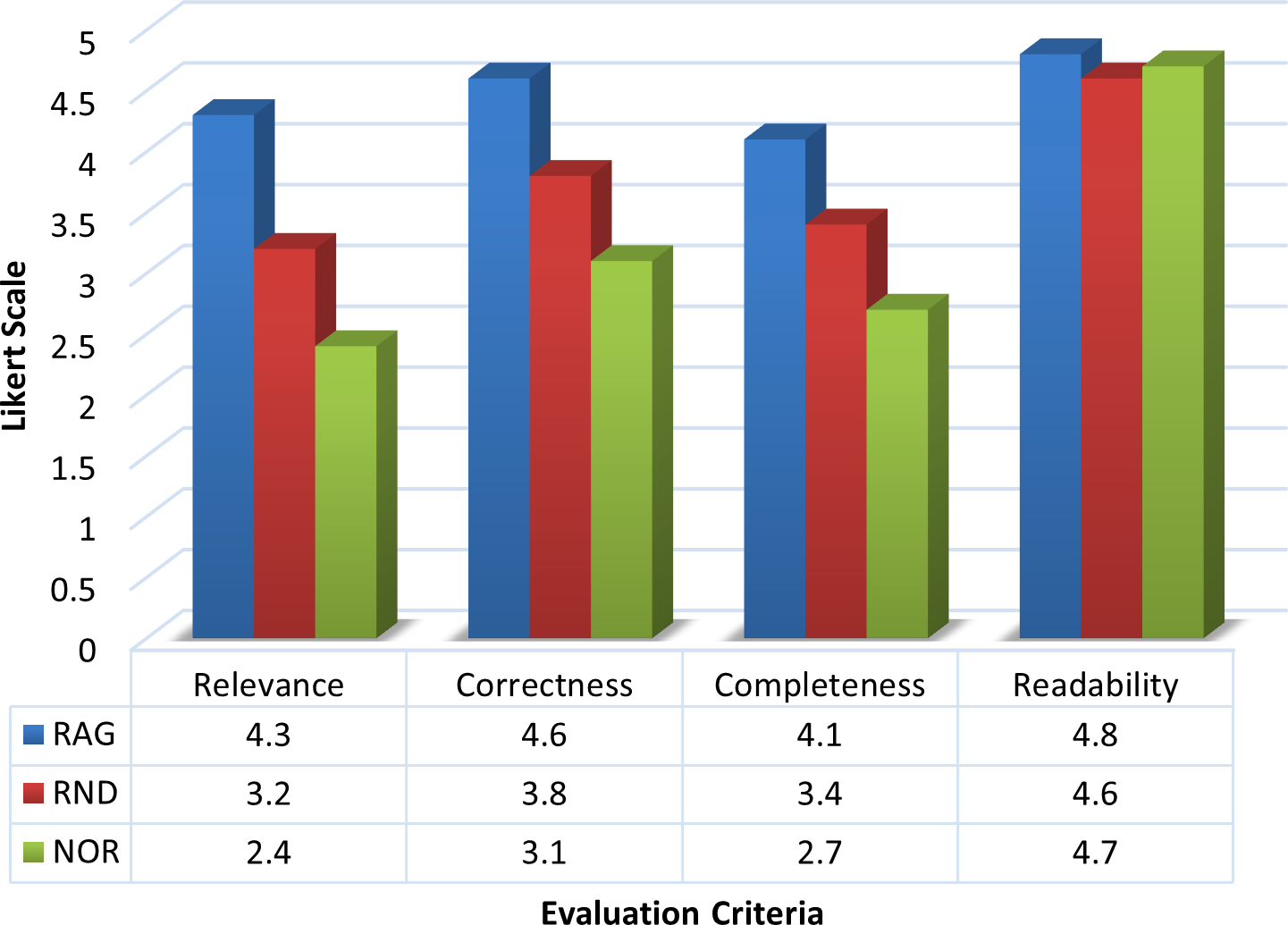}
   \caption{Human evaluation scores comparing retrieval modes. RAG outperformed both random chunk retrieval (RND) and no-retrieval (NOR) baselines in all categories.}
   \label{fig:retrieval-comparison}
\end{figure}

\begin{figure}
   \centering
   \includegraphics[width=0.95\linewidth]{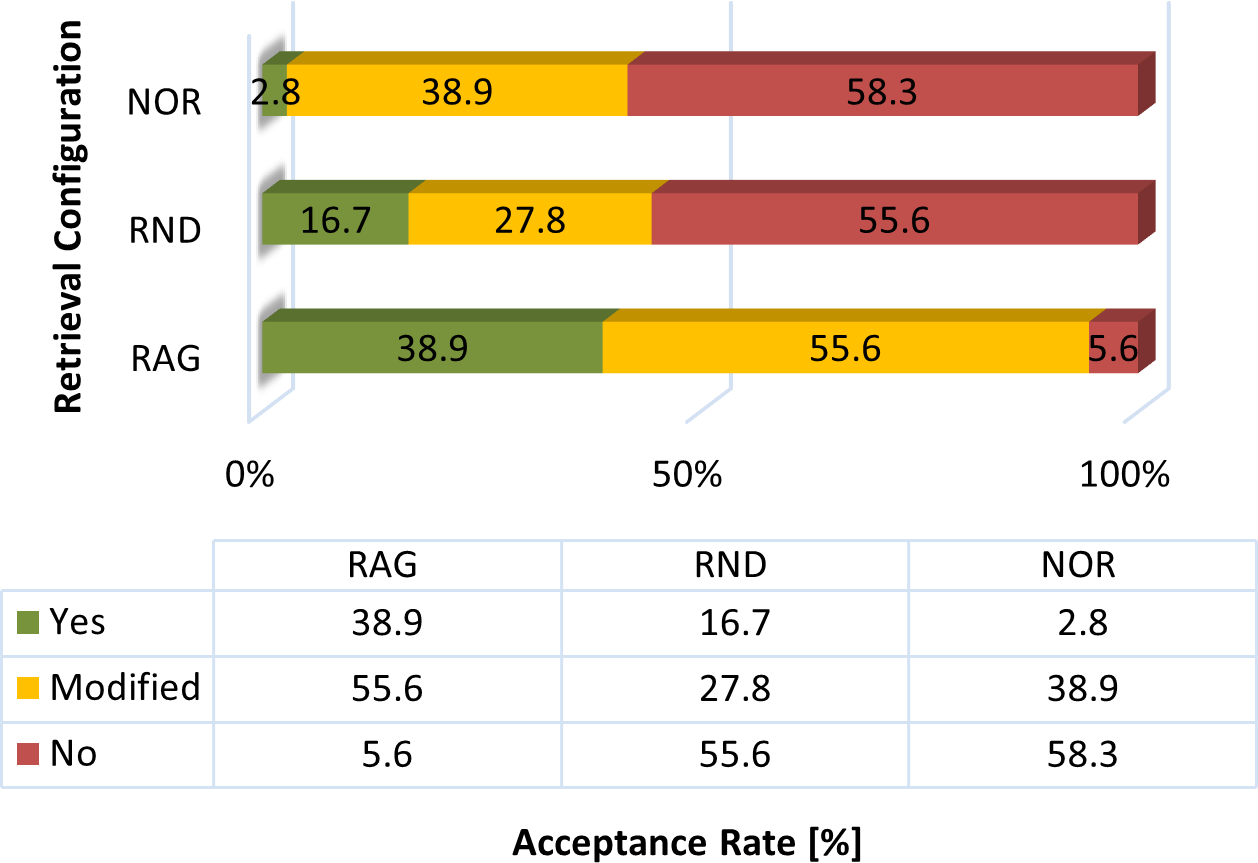}
   \caption{Acceptance distribution by retrieval mode showing that a properly designed RAG system keeps the rejection rate low.}
   \label{fig:retrieval-acceptance}
\end{figure}

From the results in Figures ~\ref{fig:retrieval-comparison} and ~\ref{fig:retrieval-acceptance}, it can be clearly seen that RAG has superior performance compared to both random retrieval and no-retrieval configurations. In the configuration that reached the highest overall human evaluation score, 38.9\% of the tests were directly accepted, 55.6\% needed some modifications and only 5.6\% of the tests had to be rewritten.

\begin{figure}[H]
   \centering
     \includegraphics[width=0.95\linewidth]{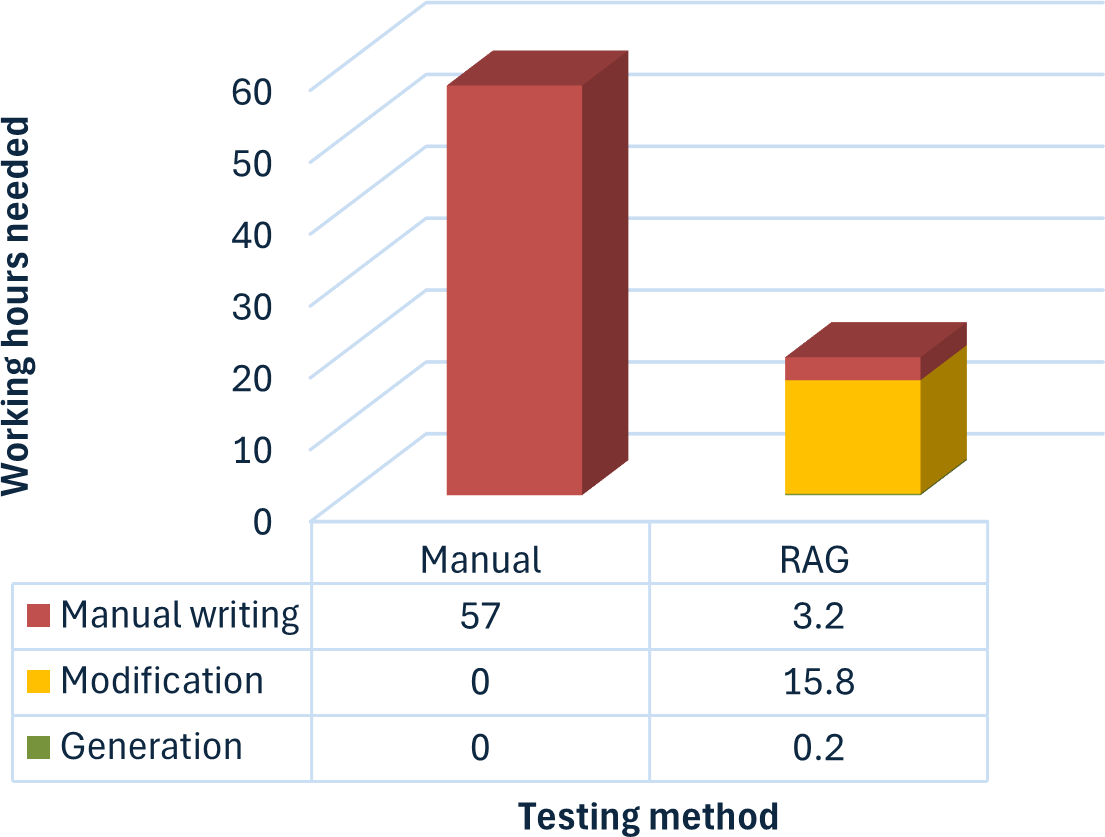}
   \caption{Time effort comparison between manual test creation and RAG-based test generation for 57 software requirements.}
   \label{fig:cost-benefit}
\end{figure}

According to the timing analysis, the speed at which the system can generate automated tests was approximately 270 tests per hour. In the testing framework used at Hydac Software, manual creation of automated tests achieves an efficiency of around 1 test per hour, as they require careful structuring and documentation within the framework. With RAG-based generation, the resulting tests are already adapted to the framework conventions resulting in little effort needed for integration with the framework. Taking into account the human evaluation and the estimated average time needed to correct tests that require modifications, the testing effort for a library with 57 software requirements could be reduced from 57 hours to 19.2 hours, constituting a time saving of 66\% (Figure~\ref{fig:cost-benefit}).

\section{Discussion and Industrial Application}

The results obtained confirm that RAG significantly improves LLM-based test generation for embedded C and thus reducing the time needed to create software tests. A key advantage of the RAG approach is that the system is adapted to the specific industrial environment. Having naming conventions, existing test suites and project documentation in the knowledge base, the generated tests are directly compatible with the testing framework with little to no need of adaptation.

Beyond immediate productivity gains, the adoption of such a system has the potential to reshape the testing workflow. Applying Christensen's idea from "Jobs to Be Done" \cite{Christensen2016}, 
the primary objective for engineers is not to write test cases, but rather to verify software behavior and ensure overall product quality. Writing tests is a necessary but repetitive activity that consumes time and cognitive capacity, increasing the risk of oversight. By filling the gap in the HYDAC Software ecosystem, the RAG tool automates the preparation of draft test cases. Crucially, because these generated tests are subsequently verified by the human engineers and existing AI test review tool, the ecosystem scales not just the number of tests, but actual confidence in the system. The test object is verified more profoundly which permits domain experts to focus on more demanding tasks such as edge-case analysis and quality assurance.

The perspective of disruptive innovation \cite{Christensen1997} suggests that new, innovative solutions do not need to match current quality across every dimension to be valuable. Instead, they must be “good enough” while delivering substantial gains in speed and scalability. At a generation speed of 270 tests per hour, with 94.4\% of the tests being directly usable or requiring only minor modifications, this threshold appears to be reached. Further refinement of the tool, such as adding a feedback loop, investigating the best configuration of each component or incorporating the newest advances in LLM technology, is expected to accelerate testing even further.

For industrial adoption, the system is most effective in organizations that maintain well-documented codebases with unified naming conventions and a rich database of legacy test suites that can be leveraged in retrieval. It is highly recommended to follow a Human-in-the-Loop (HITL) approach by using RAG as a tool to create draft tests that are then reviewed by the engineers, refined, and adjusted. This alone saves a significant amount of time and shifts expert effort from writing tests to reviewing and improving existing tests. Based on the findings of this research, a practical starting point includes fixed-size chunking, a code-optimized embedding model, hybrid retrieval with k = 5 and a knowledge base that provides the model with context on how the algorithms are implemented. With such a setup, especially when acting as the engine of a broader AI-assisted testing ecosystem, software testing productivity can be drastically increased.

\section{Conclusion}

This work presented a RAG pipeline for automated unit test generation in the context of embedded C, a domain that remains underrepresented in modern AI-driven test generation research. The results: 100\% syntactic correctness, 85\% runtime validation, and a 94.4\% test usability rate for the highest-scored configuration on the Likert scale demonstrate the significant potential of this approach. Such pre-generated tests may significantly accelerate the embedded development lifecycle and reduce the initial manual effort required from test engineers.

The impact on productivity is substantial. Traditionally, software testers required approximately 1 hour to write a single test for this specific framework. Now, by reducing that time by 66\%, the proposed system generates in minutes what would otherwise take days or weeks. More importantly, by acting as the final missing piece in the HYDAC Software toolchain, the RAG generator bridges the gap between automated project setup and automated test review. 

This complete pipeline enables test engineers to switch their attention to higher-value activities, such as designing critical test scenarios, improving coverage, and performing in-depth edge-case analysis. Achieving these results means that the paper is not only a research contribution but also a working tool that can be used in the industrial workflow, delivering measurable benefits from day one.

\bibliographystyle{IEEEtran} 
\bibliography{bibtex/literature}             

\appendix
\end{document}